\documentclass[conference]{IEEEtran}
\IEEEoverridecommandlockouts

\usepackage{cite}
\usepackage{amsmath,amssymb,amsfonts}
\usepackage{algorithmic}
\usepackage{graphicx}
\usepackage{textcomp}
\usepackage{xcolor}
\def\BibTeX{{\rm B\kern-.05em{\sc i\kern-.025em b}\kern-.08em
    T\kern-.1667em\lower.7ex\hbox{E}\kern-.125emX}}

\usepackage{hyperref}

\usepackage{booktabs}

\begin{document}

\title{Comparing Lexical and Semantic Vector Search Methods When Classifying Medical Documents
}

\author{
    \IEEEauthorblockN{Lee Harris}
    \IEEEauthorblockA{
        The University of Kent, Canterbury, UK\\
        \& \\
        TMLEP Research, Ashford, UK\\
        lah46@kent.ac.uk
    }
}

\graphicspath{{figures/}} %

\maketitle

\begin{abstract}
	
Classification is a common AI problem, and vector search is a typical solution. 
This transforms a given body of text into a numerical representation, known as an embedding, and modern improvements to vector search focus on optimising speed and predictive accuracy. %
This is often achieved through neural methods that aim to learn language semantics.
However, our results suggest that these are not always the best solution.
Our task was to classify rigidly-structured medical documents according to their content, and we found that using off-the-shelf semantic vector search produced slightly worse predictive accuracy than creating a bespoke lexical vector search model, and that it required significantly more time to execute. 
These findings suggest that traditional methods deserve to be contenders in the information retrieval toolkit, despite the prevalence and success of neural models.

\end{abstract}

\begin{IEEEkeywords}
	medical documents, automatic indexing,  semantic search, large language models

\end{IEEEkeywords}

\section{Introduction}

\cite{afantenos2005summarization} states that the internet contains a significant number of unstructured (and hence, often unusable) medical documents, and that modern technology will lead to even more becoming available in the future. 
Matching document terms against an explicit vocabulary (i.e., controlled dictionary or wordlist) is a well-established solution to the document classification (i.e., Automatic Indexing \cite{maron1961automatic}) problem, but as \cite{sebastiani1999tutorial} and \cite{jayavadivel2024historical} highlight, using humans to manually create a vocabulary may be costly and error prone. 
Alternatively, a vocabulary can be created automatically by splitting the text into blocks (i.e., tokens) that resemble words, but this may be notoriously difficult \cite{mielke2021between},
and \cite{ram2022you} highlights that this may not generalise well.
A modern approach is to feed as much of each document as possible to an AI model as training data, and to use that for future predictions. 
\cite{lai2015recurrent} introduced the Recurrent Convolutional Neural Network (RCNN) as a way to automatically learn word representations from text, and \cite{mohawesh2024fake} identified fake product reviews by combining this with the BERT \cite{devlin2018bert} large language model. 
However, these state-of-the-art models often require access to significant amounts of specific hardware (e.g., GPUs and VRAM), and neural approaches are often seen as uninterpretable, unexplainable, or otherwise too complex \cite{rudin2019stop}.
This often prevents their use alongside sensitive data.

The vector space is fundamental to AI. %
Vector search classifies each document by representing its content as an embedding (i.e., vector), and using measures and metrics in vector space to identify the most similar embedding (and its respective information) to a given (and embedded) query.
Embeddings are fundamental components in many areas of ML and AI, including search \cite{gormley2015elasticsearch} and retrieval augmented generation (RAG) \cite{lewis2020retrieval}. 
Search speed has improved significantly due to technological advancements (such as the use of GPUs \cite{jeon2021deep} and in-memory databases \cite{polychroniou2015rethinking}) and approximation techniques (ANNOY \cite{annoy} and HNSW \cite{malkov2018efficient}).  
Predictive performance has most noticeably improved due to the use of semantic content rather than matching against literal terms (respectively known as semantic and lexical vector search), and this has been made possible due to %
language models.

Our task was to split a series of rigidly-structured medical documents into 7 classes.
We extracted, preprocessed, and embedded the text in each document, and then identified the class of a new document as the same as the embedding that it was closest to. %
This process was repeated for 7 off-the-shelf embedding models. 
Lexical vector search was performed on embeddings produced by the Term Frequency (TF), Term Frequency-Inverse Document Frequency (TF-IDF), and  the Best-Matching 25 (BM25) algorithms, %
and embeddings produced by a generic word2vec(tor) \cite{mikolov2013efficient} model, a word2vec (which we refer to throughout this research as med2vec) model that was trained on medical text \cite{eyre2022launching}, a small language model which excels at quickly embedding text sentences called MiniLLM \cite{gu2024minillm}, and the state-of-the-art (as of March, 2024) mxbai \cite{embed2024mxbai} large language model were used to perform semantic vector search.

We found that lexical vector search was able to classify the medical documents that we had access to better than semantic vector search, that an unbalanced class distribution biases the predictive accuracy produced by vector search, and that the amount of data positively correlates with the predictive accuracy produced by vector search.
These suggest that it may not be necessary to use an off-the-shelf semantic vector search model (even though these are amongst the most successful and popular) as producing a bespoke lexical vector search model may be faster and more predictive.
We are not the first researchers to investigate the vector space as a way to classify medical documents, and recent research by \cite{rohanian2024rapid} has produced similar findings.
Additionally, AI technology is advancing incredibly quickly, and our results highlight the current technological capabilities and limitations.

\section{Background}
\label{SEC:background}

\subsection{Vector Search}
The goal of vector search is to identify the stored embeddings that are most relevant to a given or created embedding.
The relevance is calculated by the lazily-learnt k Nearest Neighbour (kNN) algorithm. %
This predicts the class of a new data example, to be the same as the ones with the maximum inverse cumulative distance or similarity, as defined by the sum of the k closest data examples.
The closest embedding would be chosen if k were set to its smallest possible value, 1.
Popular distance and similarity measures include the L1 and L2 norm (i.e., respectively, Manhattan and Euclidean) distance, infinity-norm distance, cosine similarity, Mahalanobis distance, and the hamming difference \cite{xie2023brief}.
We use the Euclidean distance in this research, and this is defined as 

\begin{equation}
	D(a,b)=\sqrt{\sum_{k=0}(b_k-a_k)^2}
\end{equation}

\noindent where $a$ and $b$ are vectors of the real numbers, $k$ is a vector dimension, and $D$ is the distance.

Lexical searches identify the suitability between an embedding and a query by using key words and terms. %
This can be problematic if the query is search engine-esque, as it may be  uncommon for the most useful or relevant embeddding to a query to contain the exact query words, but this simple logic often leads to searches that are very fast.
Examples of lexical search algorithms are Term Frequency (TF; Sec.~\ref{SSEC:tf}), Term Frequency $\times$ Inverse-Document-Frequency (TF-IDF; Sec.~\ref{SSEC:tf-idf}), and Best-Matching 25 (BM25; Sec.~\ref{SSEC:bm25}). 

Semantic search is often seen as an improvement on lexical search.
This uses contextual knowledge (often inferred from data) to capture semantic information from a query, and match it to an embedding. 
This can be a slow process, but semantic embeddings have been shown to achieve  higher predictive accuracy than lexical embeddings \cite{yu2014improving}.
However, it is because of the knowledge inferred from the training data that these models are accused of creating falsehoods (i.e., hallucinations \cite{perkovic2024hallucinations}).
The most popular method of creating semantic embeddings is using neural networks (SEC.~\ref{SSEC:neural_methods}), and examples are GloVE \cite{pennington2014glove}, word2vec (Sec.~\ref{SSSEC:word2vec}), and language models (Sec.~\ref{SSSEC:language_models}).

The choice of whether to use lexical or semantic search is often motivated by the type of output that is desired.
Lexical search produces sparse embeddings that contain mostly 0, and a few useful numbers, where as semantic embeddings are densely packed, and it is common for these to be filled entirely with unique values.
This is the synonymous to disentanglement \cite{carbonneau2022measuring} and superposition \cite{elhage2022toy} discussion in the domain of information encoding.
Computational resources may be wasted by a sparse matrix, but the meaning of altering the values in one of these dimensions is clear, and it would not impact on the other dimensions.

\subsection{Term Frequency (TF)}
\label{SSEC:tf}
The simple Term Frequency (TF) method generates embeddings (i.e., matrices) whose value is the number of times that each token (i.e., word or term) appears in each body of text (e.g., a document).
This is formally written as 

\begin{equation}
	TF(t, w) = f(t, w),
\end{equation}

\noindent{}%
where $w$ is a word, $t$ is a body of text, and $f$ is a function that counts the number of times that $w$ appears in $t$.
A common variation of this is to divide each by the total number of words in each body of text \cite{lv2011lower}, and other researchers continue to explore method variations and improvements \cite{azam2012comparison}.

\subsection{Term Frequency * Inverse Document Frequency (TF-IDF)}
\label{SSEC:tf-idf}
The TF measure may undesirably assign importance to (stop) words which do not reveal useful text data.
One way to avoid this problem is by introducing the popular Inverse Document Frequency (IDF) heuristic. 
This measures how important a term is by how commonly it appears across all of the texts in a corpus.
The motivation is that rare words should have higher importance. 
IDF is formally written as

\begin{equation}
	IDF(w, c) = \log{}\bigg(\frac{|c|}{c_w}\bigg),
\end{equation}

\noindent{}%
where $w$ is a word, $c$ is a text corpus, $|c|$ is the size of the $c$, and $c_w$ indicates how many texts in $c$ contain $w$.
A small numeric constant is sometimes added to the denominator to avoid division by 0.
The new TF-IDF method is the product of the TF and the IDF, and this written as 

\begin{equation}
	TF-IDF(w, c) = \sum_{t \in c}{(TF(c_t, w) \times IDF(w, c))}
\end{equation}

\noindent{}%
where $w$ is a word, $c$ is a corpus, and $c_t$ is a corpus text.

\subsection{Best Match 25 (BM25)}
\label{SSEC:bm25}
The strength of the embeddings produced by TF-IDF may be disproportionate between documents as there is no threshold of how important a word can be, and a word may appear numerous times in a piece of text, but the text that it appears in may be larger than the cumulative text of all other corpus texts.
These issues were addressed by the Best Match 25 (BM25) algorithm \cite{robertson1995okapi}.
The BM25 formula calculates a score for each corpus text that represents its relevance to a given query, with the most relevant documents receiving the highest score.
We consider fixed-size embeddings in this research, and we set the embedding size to the length of the corpus. 
This was one of the most commonly used embedding methods in search engines prior to the introduction of neural models, and it is still used regularly in some modern search tools (such as Elastic search \cite{gormley2015elasticsearch}) due to its speed.
We use a variation of BM25 known as BM25+ \cite{lv2011lower} in this research. 
This introduces a lower bound to the term frequency normalisation that prevents texts  which do not contain a query term from being incorrectly assigned large importance.

This algorithm introduces two hyperparameters, $k_1$ and $b$, which respectively control how quickly the embedding value saturates, and the importance of document length. 
$\delta$ is introduced by BM25+. 
These respectively have default values of $1$, $0.75$, and $1$, and the BM25 formula (with addition of $\delta$) is written as

\begin{multline}
	BM25(w,c) = \\
	\sum_{t \in c} IDF(w, c) \times \bigg(\frac{TF(t, w) \times (k_1 + 1)}{TF(t, w) + k_1 \times (1 - b + b \times \frac{|t|}{avgtl}} + \delta \bigg),
\end{multline}

\noindent{}%
where $c$ is a corpus, $w$ is a word or query, $t$ is text in $c$, $|t|$ is the text length, and $avgtl$ is the average text length in $c$.

\subsection{Neural Methods}
\label{SSEC:neural_methods}
The use of neural networks and deep learning has grown significantly in recent years. 
These map a query, through a series of intermediate layers, to an output. 
The learnt knowledge is then stored in the model's parameters (i.e., weights and biases).
These models are able to handle a variety of input formats \cite{yu2019review}, and language models have performed exceptionally well on a wide range of tasks \cite{liu2023summary}.

\subsubsection{Word2Vec}
\label{SSSEC:word2vec}
These techniques convert a large corpus of text into a vector space containing word representations, and words with similar meanings have similar embeddings. 
The model works by considering words in pairs, where one word acts as 
the input and the other as the "output", and the goal is to predict the output word given the input word (or vice versa).

\subsubsection{Language Models}
\label{SSSEC:language_models}
A Large Language Model (LLM) is able to generate text by analysing patterns in large text corpuses, and using this information to generate probability distributions that are able to predict the next word that follows a given input. 
These currently achieve the highest predictive performance, although it has recently been shown that (small) language models which have been trained to solve a specific task outperform general LLMs \cite{schick2020s}.
The minilm and mxbai model in our research are built on top of the BERT \cite{devlin2018bert} LLM.

\section{Methodology}
\subsection{The Data}
We extracted 1472 documents consisting of various personal details (names, addresses, etc.) and medical histories (ailments, prescriptions, medications, appointments, notes, etc.)  from 100 medical cases that were collected over a 4-month window. %
These contained various images, such as logos and signatures, and the text in each page was written from left to right on a vertically-oriented white background. %

\subsection{The Task}
Our task was to assign one of 7 classes to each medical document.
The names of the classes have been replaced throughout this research with letters of the alphabet, with the default being E.
These were set according to 1) explicit filename content that was assigned at data collection, and 2) obvious visual artefacts. 
Unfortunately, these were often missing (e.g., because they were unique, ambiguous, etc.) in which case the document class became the default.

The classes shown in this research are the result of anonymisation, and the medical data has not been made available due to the sensitive nature of this.

\subsection{Document Transcription}
The documents were written in an established typeface using a common document preprocessor (e.g., Microsoft Word), and these were stored in either portable (i.e., PDF) or .docx format. 
The text was extracted from documents with a .docx extension using the mammoth\footnote{\url{https://pypi.org/project/mammoth/}} python package, and it was extracted from the bytes in the text layer of a PDF document using the pypdfium2\footnote{\url{https://pypi.org/project/pypdfium2/}} python package.
The Tesseract \cite{smith2007overview} Optical Character Recognition (OCR) tool was used to extract a digital text transcription if a PDF's textual layer was empty or corrupt (for instance, if it contained scanned copies of paper documents), but this was often slower (and potentially less accurate) than using the explicit byte information. %

\subsection{Text Preprocessing}
This 
is an important step when handling text that may contain inconsistencies, noise, or irrelevant data, especially when performing vector search \cite{rahimi2023impact}. 
First, we transformed each transcribed text to lowercase.
Next, we removed all punctuation, numbers, dates, parentheses, and unicode characters by choosing to only retain spaces, and letters between a and z. 
We then tokenised each document, effectively creating words, by splitting on the space characters. 
Next, we performed stemming to convert each word into its base form. %
The community-run snowball\footnote{\url{https://snowballstem.org/}} project was used via a python decorator called PyStemmer\footnote{\url{https://pypi.org/project/PyStemmer/}}.
Finally, stop words (e.g., and, in, of, the, etc.) defined by the Natural Language ToolKit\footnote{\url{https://pypi.org/project/nltk/}} (NLTK) were removed from the text.

\subsection{Embeddings} %
We converted each document's transcribed text into 7 embeddings using the 7 embedding algorithms and models described in Sec.~\ref{SEC:background}: 1) TF, 2) TF-IDF, 3) BM25, 4) word2vec, 5) med2vec, 6) minillm, and 7) mxbai.
The sklearn\footnote{\url{https://scikit-learn.org}} implementations of TF and TF-IDF were used, the bm25s\footnote{\url{https://github.com/xhluca/bm25s}} \cite{lu2024bm25s} python package's fast implementation of the default BM25 algorithm was used, the SpaCy\footnote{\url{https://spacy.io/}} and MedSpaCy\footnote{\url{https://github.com/medspacy/medspacy}} \cite{eyre2022launching} implementations of word2vec and med2vec were used, and ollama\footnote{\url{https://ollama.com/}} \cite{dubey2024llama} implementations of minillm and mxbai were used.
We used the largest parameter variants of each model.

A vocabulary was created for the embedding methods that needed it (i.e., TF, TF-IDF, and BM25), and this was an alphabetically-sorted set of all words from all documents in the respective training set.

\subsection{Vector Search}
We used chroma\footnote{\url{https://github.com/chroma-core/chroma}} to store the embeddings, and to perform nearest-neighbour vector search on these. 
By default, Chroma performs approximate search, but we disabled this by setting the 
\textit{hnsw:search\_ef} hyperparameter to a much larger value than the default (10 vs 50). 
This defines the trade-off between recall and computational cost. %
We used the L2 norm (i.e., Euclidean) distance metric. %

\subsection{Default Hyperparameters}
Our default approach was to store/train on 75\% of the embeddings, and to test against the remaining 25\%. 
The default number of nearest neighbours was 5, under the initial assumption that more is better, and each experiment was repeated 10 times: each time shuffling the data with a different random seed. 
Any of the default parameters that changed between experiment (such as the number of nearest neighbours) are explicitly indicated in the respective figure captions. 
The embedding method names appear in lowercase.

\subsection{Ethical Considerations and Compliance} 
The classes are for internal-company use only, and these will not be used to produce decisions or actions.
There are no consequences to any data stakeholders if the document classification were to perform poorly or incorrectly. 
The original documents are unaltered by the classification process.

All data was stored on the company's cloud storage
and viewed and manipulated by humans and machines exclusively on the company premises by those who were authorised to do so.
The company rigidly adheres to the ISO 12007 standard, and the secure storage of data.

\section{Results}

This section presents the results of the various classification experiments.
RGB bar and line colours (rather than greyscale tones) are used consistently in each figure throughout this section to distinguish amongst the many embedding methods.

\begin{figure}[ht]
	\begin{tabular}{cccc}
		\multicolumn{4}{c}{T-SNE Visualisations} \\
		\multicolumn{4}{c}{75:25 Data Split} \\
		legend & tf & tf-idf & bm25 \\
		\includegraphics[trim={0 0 0 0},clip,height=1.8cm,width=.08\linewidth{}]{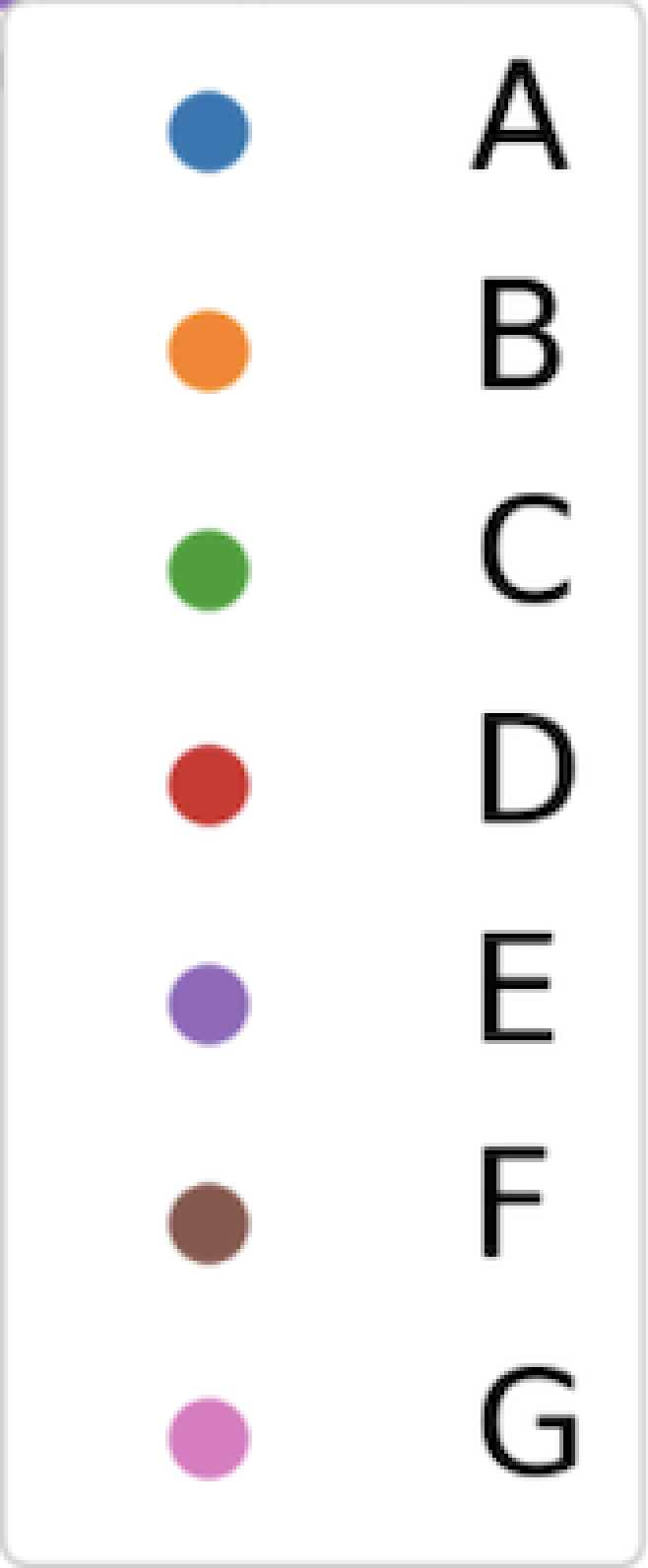} &
		\includegraphics[trim={0 0 0 0},clip,width=.21\linewidth{}]{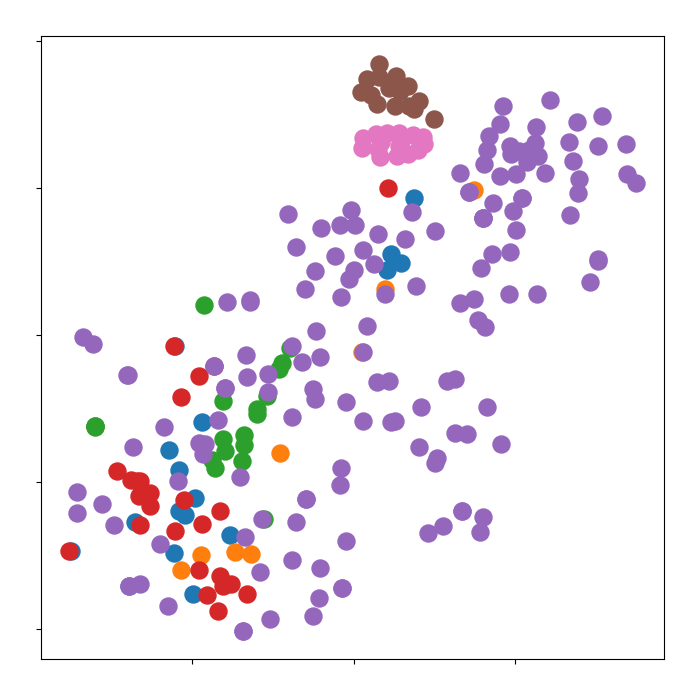} & 
		\includegraphics[trim={0 0 0 0},clip,width=.21\linewidth{}]{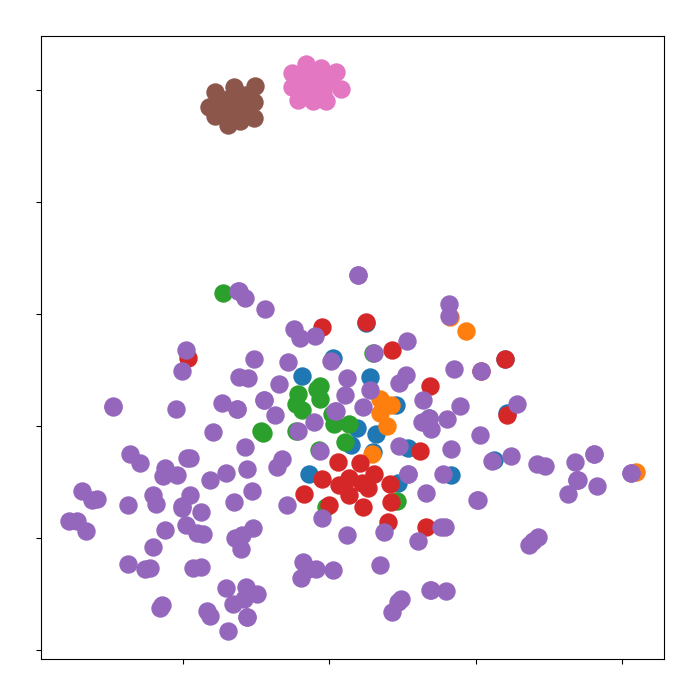} & 
		\includegraphics[trim={0 0 0 0},clip,width=.21\linewidth{}]{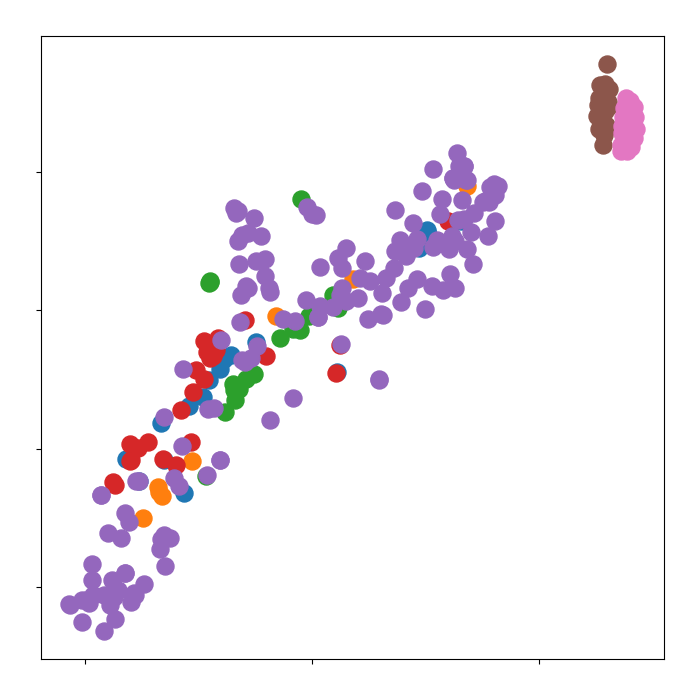} \\
		
		\includegraphics[trim={0 0 0 0},clip,width=.21\linewidth{}]{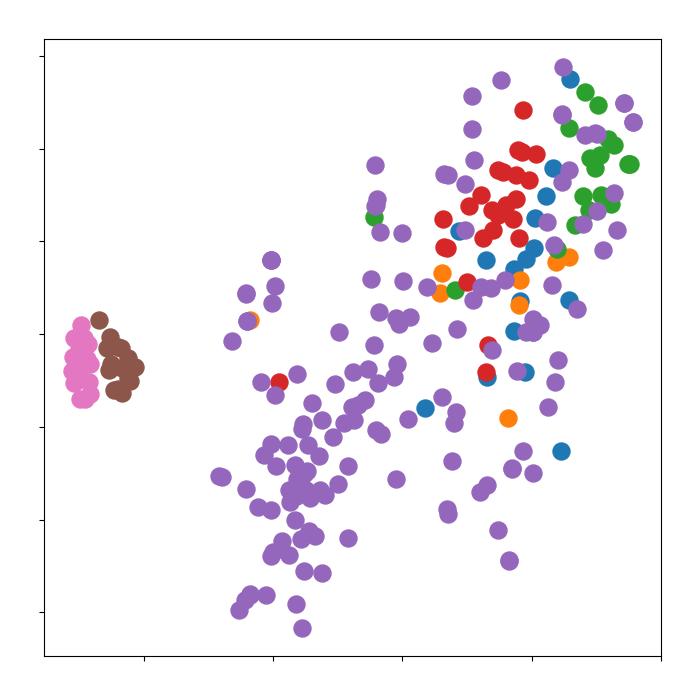} & 
		\includegraphics[trim={0 0 0 0},clip,width=.21\linewidth{}]{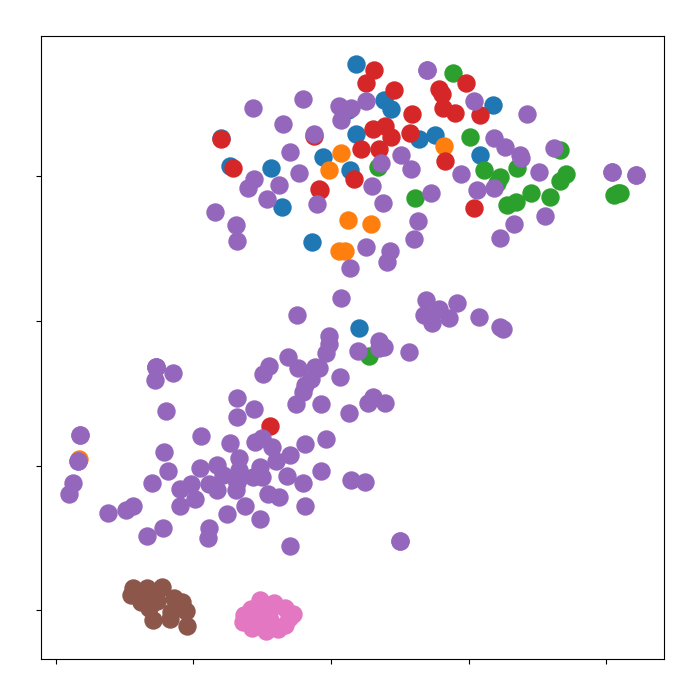} &
		\includegraphics[trim={0 0 0 0},clip,width=.21\linewidth{}]{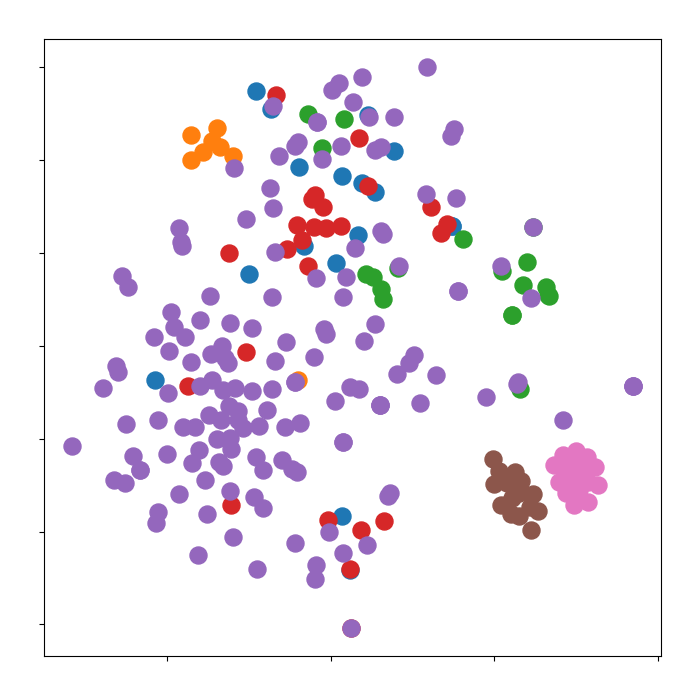} & 
	
		\includegraphics[trim={0 0 0 0},clip,width=.21\linewidth{}]{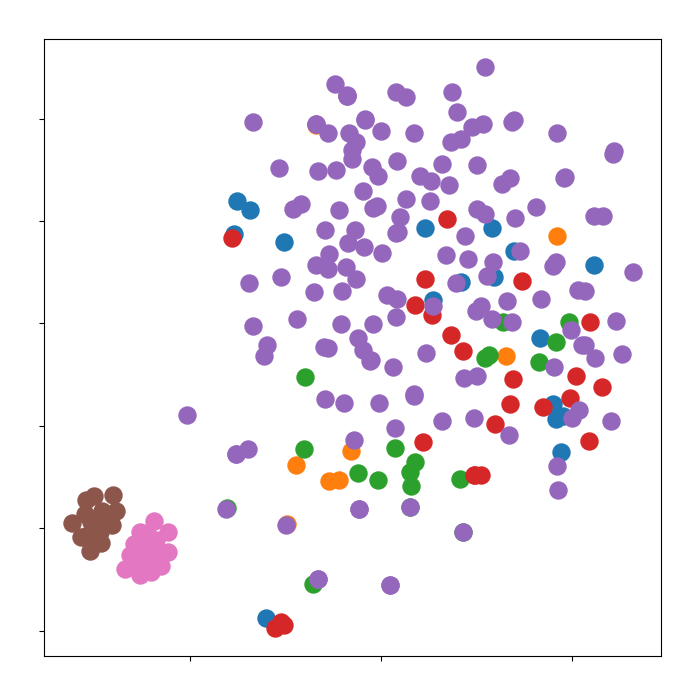} \\
	Word2Vec & Med2Vec & MiniLM & Mxbai \\
	\end{tabular}

\caption{These 7 scatterplots show how the T-Stochastic Neighbourhood Embeddings (T-SNE) algorithm compresses 250 embeddings produced by each embedding method explored in this research. 
The legend is shown in the top left.
Empirically, these are typical examples of what each embedding method would produce from a 75\%:25\%/stored:test data split, although they show the first 250 data examples from a single dataset that was randomly shuffled using a random seed of 0.
The values on the X and Y axis are ignored, and what is significant is each embedding's proximity to the other embeddings. 
The point colours correspond to the different classes.
In each plot, the embeddings that are labelled using classes F and G (respectively, brown and pink) are very distinct from the others, and there are many more class E embeddings than there are for other classes.
}

\label{FIG:t-sne_visualisations}
\end{figure}

Tab.~\ref{TAB:class_statistics} describes the amount of words in the document texts before and after preprocessing, organised by each document class. 
The number and spread of the words in each category change consistently before and after the texts are processed, and there is a clear division between the amount of words in the different categories.  %
Significantly more documents belong to the default class (E) than the others classes, and classes B and F are associated with the fewest documents. 
Documents in classes F and G are very rigidly structured. %
This is shown by the small standard deviation (std), and by the very small difference between quartiles 3 and 1 (i.e., 75\% and 25\%).
The documents in class C are more rigidly structured than classes A, B or D, and the structure of the documents in class E are very inconsistent.
These results agree with those shown in Fig.~\ref{FIG:t-sne_visualisations}.

\begin{table}
	\begin{tabular}{lccccccc}
	\hline
	\multicolumn{8}{c}{\textbf{Document classification}} \\
	\hline

	\hline
	\multicolumn{8}{c}{Pre Text Processing} \\
	\hline

	 & A & B & C & D & E & F & G \\
	\# & 112 & 54 & 115 & 108 & 803 & 74 & 94 \\
	mean & 12964 & 16358 & 10145 & 16441 & 16288 & 1280 & 1039 \\
	std & 8429 & 12728 & 4783 & 10044 & 30301 & 359 & 20 \\
	25\% & 6957 & 7813 & 6554 & 9508 & 2365 & 1016 & 1025 \\
	50\% & 11422 & 15746 & 9455 & 14866 & 6907 & 1204 & 1033 \\
	75\% & 17326 & 21897 & 12137 & 21322 & 15804 & 1425 & 1047 \\

	\hline
	\multicolumn{8}{c}{Post Text Processing} \\
	\hline

	 & A & B & C & D & E & F & G \\
	\# & 112 & 54 & 115 & 108 & 803 & 74 & 94 \\
	mean & 12456 & 15862 & 9828 & 15029 & 15144 & 1057 & 924 \\
	std & 8190 & 12449 & 4643 & 9458 & 28142 & 261 & 18 \\
	25\% & 6446 & 7494 & 6432 & 8164 & 2193 & 864 & 910 \\
	50\% & 10952 & 14974 & 9206 & 13458 & 6376 & 996 & 920 \\
	75\% & 16662 & 20913 & 11794 & 19708 & 14828 & 1160 & 931 \\

	\hline
	\\
	\end{tabular}

\caption{Descriptive statistics about the document text before and after preprocessing, grouped by the class. 
The table columns correspond to the classes, and the rows correspond to the document statistics, these are: %
1) the number of documents in that class, 
2) the average number of words in those documents, 
3) the standard deviations, 
4-6) respectively show the number of words in the 25\%, 50\%, 75\% document, when these are sorted by the number of words.
}

\label{TAB:class_statistics}
\end{table}

Fig.~\ref{FIG:t-sne_visualisations} shows 250 embeddings that are compressed into 2D scatterplots using the T-SNE algorithm. 
The embeddings correspond to the same data examples, and these were empirically representative of the class distribution. 
The embeddings with a target class of F or G are clearly separated from the others in every scatterplot. 
The embeddings corresponding to class C are sometimes distinct. 
The default class (E) was represented by many more embeddings than the other classes, but there were often concentrations of embeddings that corresponded to this class despite these being incredibly numerous and spread.

Tab.~\ref{TAB:embedding_dimensions} shows the number of dimensions produced by each embedding method when the data was split 75\%:25\% between the storage and test sets.
The first column shows the embedding method's name, the second column indicates the type of vector search  performed by each embedding, the third column is the number of embedding dimensions, and the final column shows the amount of memory required to store a single document embedding.
The amount of dimensions in the lexical vector search embeddings was significantly larger than those produced by the semantic vector search embeddings, and these used much more computer memory.

\begin{table}
	\begin{tabular}{|c|c|c|c|}

	\hline
 	Method & Vector Search Type & Dimensions & Memory (bytes) \\
 	\hline
    tf & lexical & 42208 & 337792 \\
    tf-idf & lexical & 42208 & 337792 \\
    bm25 & lexical & 1020 & 8288 \\
    word2vec & semantic & 300 & 2528 \\
    med2vec & semantic & 300 & 2528 \\
    minilm & semantic & 384 & 3200 \\
    mxbai & semantic & 1024 & 8320 \\
	\hline
	\end{tabular}
\\

\caption{The average test embedding properties.
75\% of the data was stored in the vector database, and 25\% was used for testing.
The lexical search embeddings are mostly filled with 0, while most (if not all) of the embedded values used in semantic search are non-zero. 
}

\label{TAB:embedding_dimensions}
\end{table}

\begin{figure}[ht]
	\begin{tabular}{c}
	Vector Search Classification Statistics \\
	75:25 Data Split, K = 5 \\
	\hline 
	~~~~~Precision \\
	\includegraphics[trim={0 1.5cm 0 0.4cm},clip,height=2cm,width=0.99\linewidth{}]{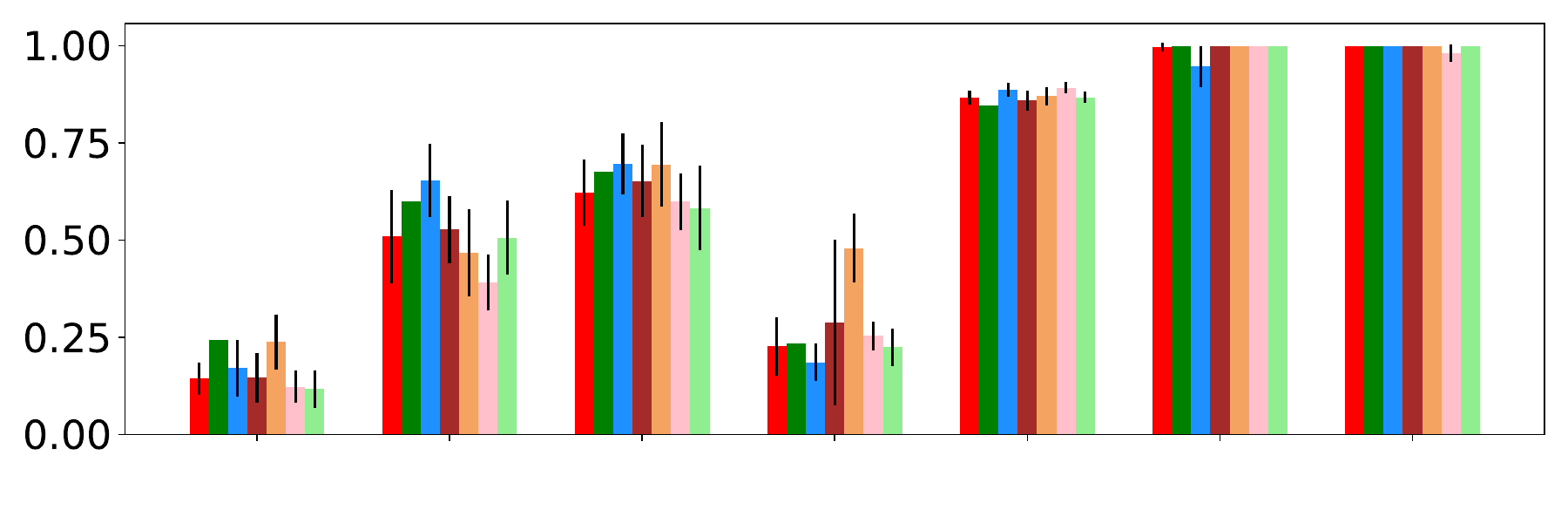}\\
	~~~~~Recall \\
	\includegraphics[trim={0 1.5cm 0 0.4cm},clip,height=2cm,width=0.99\linewidth{}]{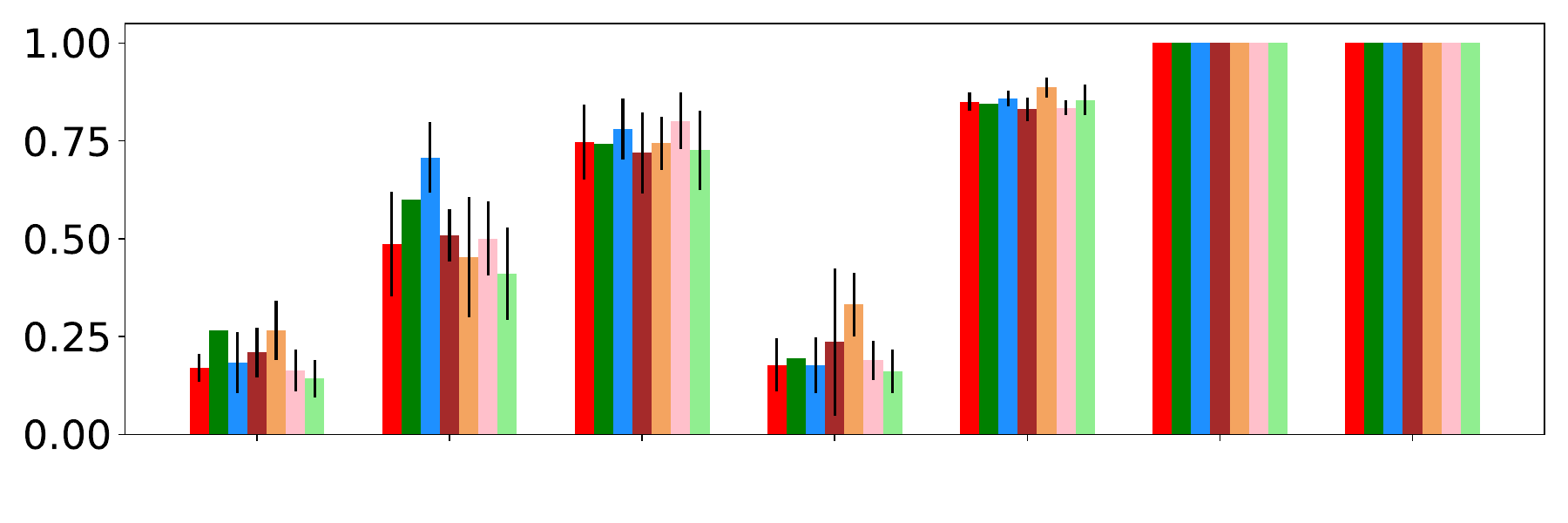} \\
	~~~~~F1 \\
	\includegraphics[trim={0 0.6cm 0 0.4cm},clip,height=2.5cm,width=0.99\linewidth{}]{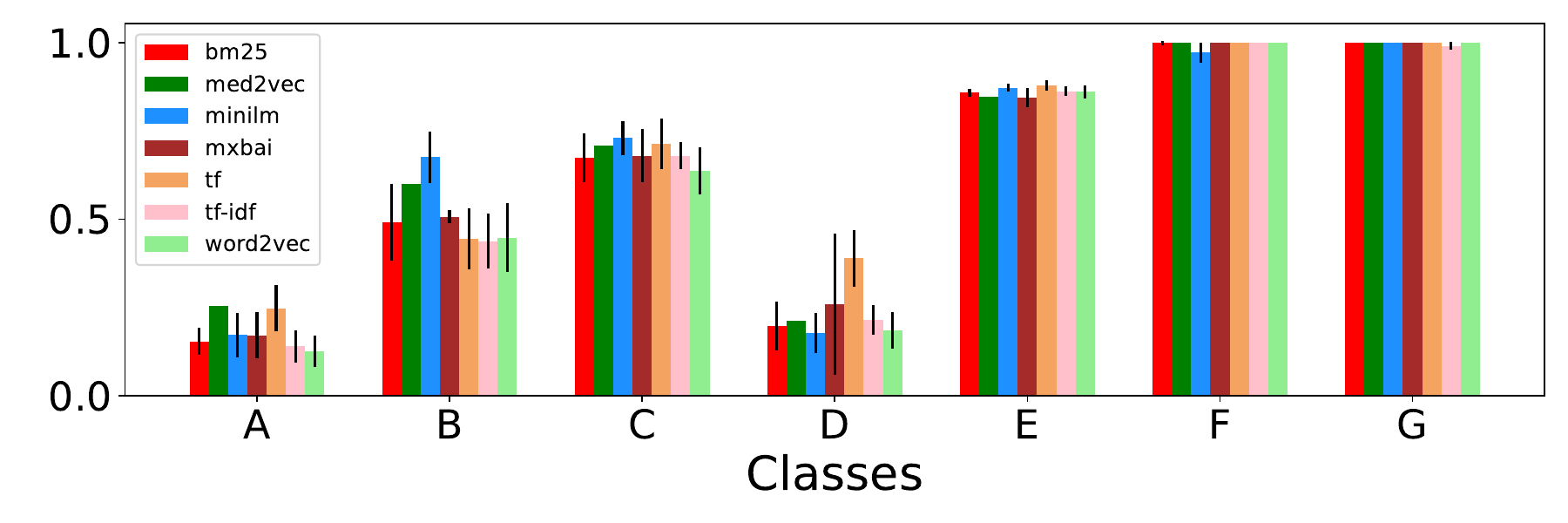} \\
	\\
	\end{tabular}

\caption{The average precisions, recalls, and f1-scores resulting from 10 classification experiments. 
75\% of the data was stored in the vector database, and the other 25\% was used as a test set. 
This experiment was then repeated with 10 different random seeds.
Vector search was performed with a k-nearest-neighbour value of 5, and the Euclidean distance metric.
The Y-axis shows values between 0 and 1, and each bar on the X-axis shows the metric produced by each embedding method, for each class. 
The bar colours indicate the embedding method that was used, and the black lines at the top of each bar show the standard deviations.
}

\label{FIG:classification_statistics}
\end{figure}

Fig.~\ref{FIG:classification_statistics} shows the mean precision, recall and F1 score that resulted from the 10 75\%:25\% stored:test classification experiments. 
The documents belonging to classes F and G were classified correctly (or very nearly) by every embedding method. 
The default document class (E) was nearly classified correctly, followed closely by documents belonging to class C. 
Documents belonging to class B were classified correctly, but documents belonging to A or D were rarely classified correctly.

\begin{figure}[ht]
    \centering
    \includegraphics[trim={0 0cm 0 0cm},clip,width=\linewidth{}]{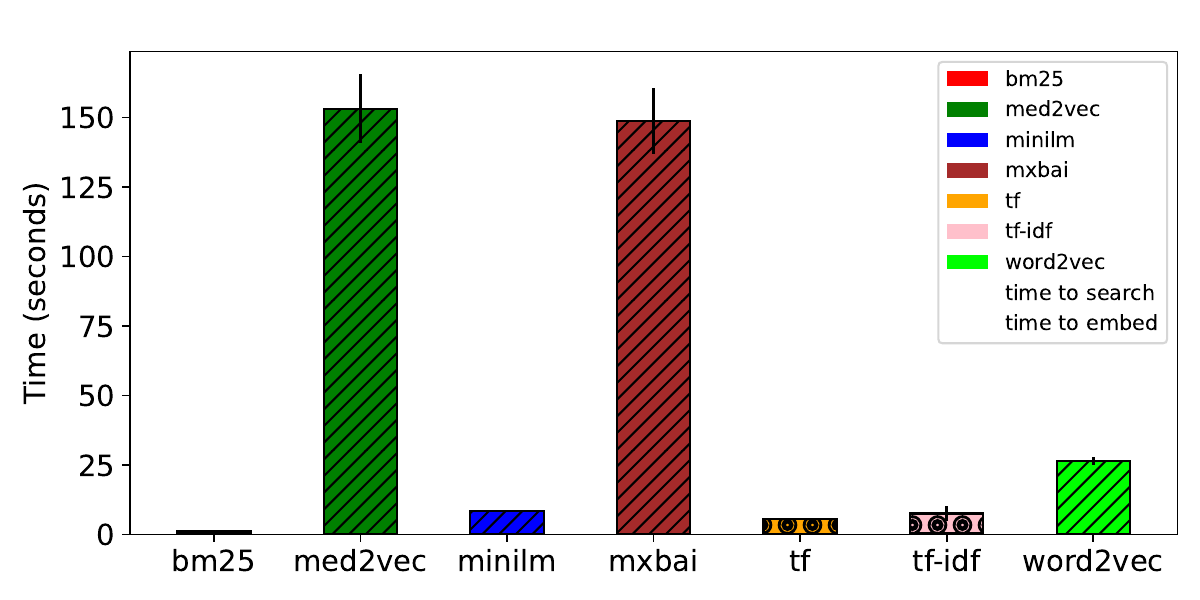}

    \caption{The time, in seconds, that each embedding method took to embed and search 25\% of the data.
    These bars do not include the time required to embed and store the other 75\% of data. 
    This process was repeated 25 times, where the data was shuffled by the n$^{th}$ random seed before recording took place.
    The bar colours indicate which embedding method was measured, while the bars prefaced with forward-slashes show how long each embedding method took to embed the transcribed  document text, and the bars prefaced with circles show how long it took to perform 5-nearest neighbour vector search using the respective embedding.
    Each embedding method had results for the time required to embed the data, and the time required to perform vector search, although these may not be visible if either time were near 0.
    The standard deviation for each bar was represented by a black line in the centre.
    The tops of each embedding method's bar indicate the total execution time.
    These experiments were run on a MacBook Pro, that featured an Apple-silicon M2 CPU, and 32GB of DDR4 RAM. No GPU was used. }
    
    \label{FIG:execution_times}
\end{figure}

Fig.~\ref{FIG:execution_times} shows how long each embedding method took to embed and search each document in the test set. %
The very simple TF and TF-IDF methods quickly embedded the document text, but vector search using the embeddings was much slower. 
By contrast, the word2vec, med2vec, and mxbai methods slowly embedded the document text, but the amount of time required to perform vector search was negligible. 
minilm, while still being a neural model, was much faster at embedding the text.
BM25 produced relatively-small embeddings very quickly, and was very quickly able to perform vector search with these.

\begin{figure}[ht]
    \centering
    \includegraphics[width=\linewidth]{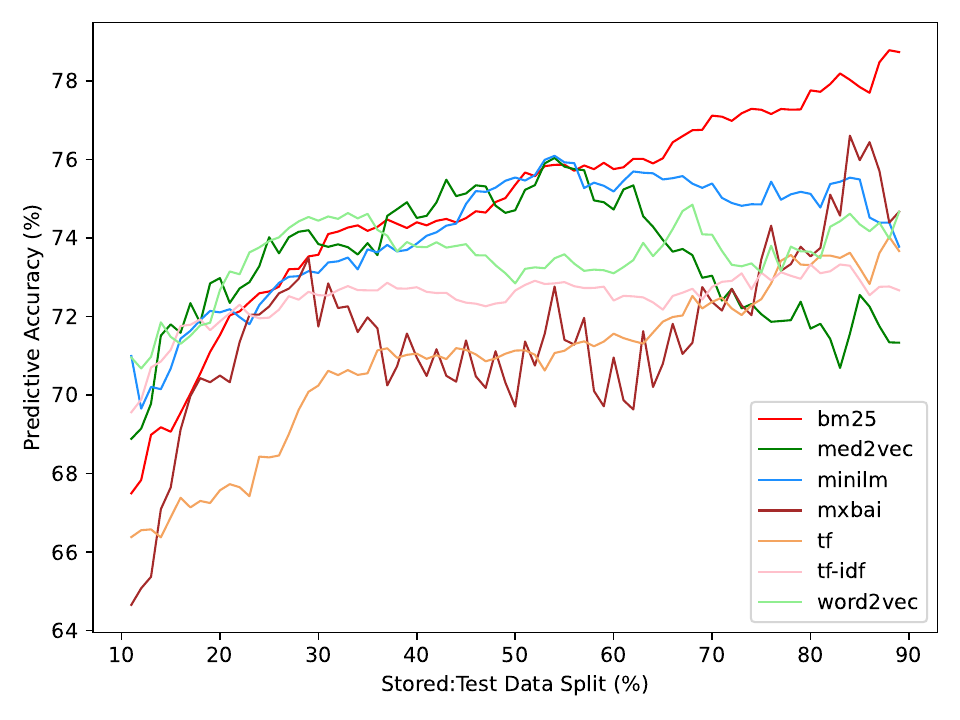}

    \caption{Each embedding method's average classification performance in response to the amount of stored and test data.
    The vector search considered 5 nearest neighbours.
    The X-axis indicates the data split between the stored set and the test set, and the Y-axis indicates the mean classification correctness.
    This was averaged over 10 repetitions, with the $n^{th}$ repetition shuffling the data using a random seed of $n$.
    }
    
    \label{FIG:data_splits_and_predictive_performance}
\end{figure}

Fig.~\ref{FIG:data_splits_and_predictive_performance} shows the average predictive performance of each embedding method between 10:90 and 90:10 stored data:test data, and this was repeated with 10 different random seeds.
BM25 and the minilm achieved the highest predictive accuracy across most data splits, while TF and the mxbai model achieved the lowest. 
The embeddings used in lexical vector search generally achieved higher predictive accuracy than the  embeddings used in semantic vector search.
The predictive accuracy of the embedding methods increased as a result of increasing the amount of stored/training data.

\begin{figure}[ht]
    \centering
    \includegraphics[trim={0 0cm 0 0cm},clip,width=\linewidth]{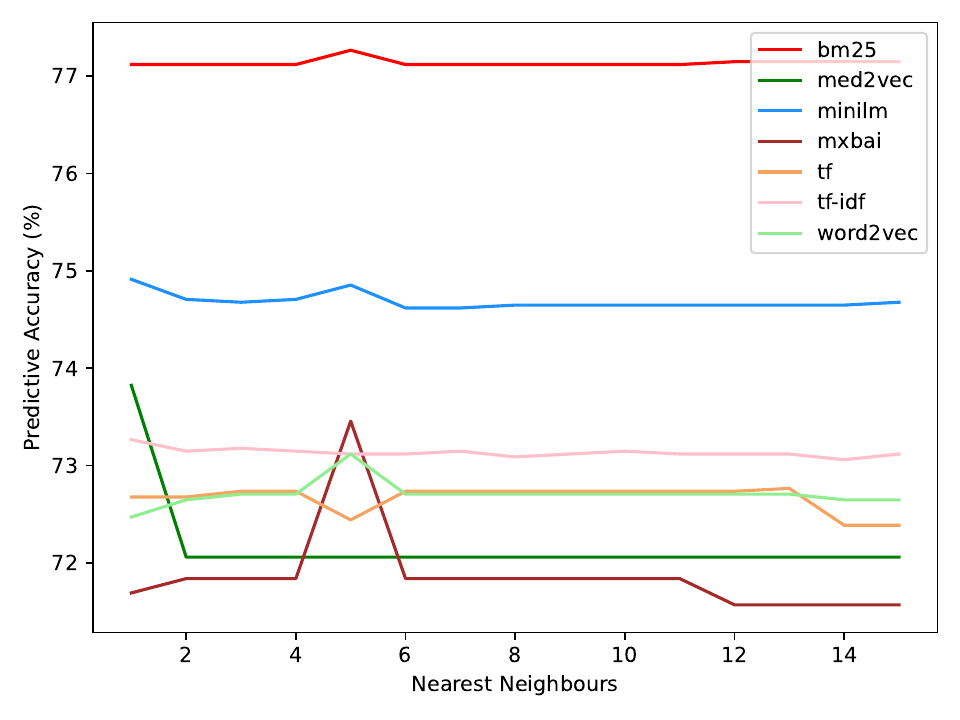}

    \caption{How the number of nearest neighbours impacts the predictive accuracy of vector search.
    The data was split 75:25, with 75\% of the data being stored, and 25\% being used for testing.
    The number of nearest neighbours is shown on the X-axis, and this increased from left-right, up to 15.
    The line colours correspond to the same embedding methods as in the other figures.  
    }
    
    \label{FIG:knn_and_predictive_performance}
\end{figure}

Fig.~\ref{FIG:knn_and_predictive_performance} shows how the average predictive accuracy at the 75\%:25\%/stored:test data split changes depending on the number of nearest neighbours used during vector search.
The predictive accuracy decreased slightly when the number of nearest neighbours was increased, and a noticeable predictive accuracy change occurred around the value 5, but changing this hyperparameter mostly had a negligible impact.

\section{Discussion}
\subsection{Bespoke data may not require complex AI}
The results reveal that fitting a custom model to bespoke data may outperform an off-of-the-shelf one, and the distance of this gap may change depending on the task.
For instance, the lexical (red, brown, and yellow) and semantic (dark green, blue, brown, and light green) vector search methods produced comparable predictive accuracy on the rigidly-structured medical data, but comparing the most predictive lexical and semantic vector search methods (respectively BM25 and minilm) reveals that BM25 embedded and searched slightly quicker than minilm, and produced higher predictive accuracy. 
Additionally, the lexical vector search methods were significantly faster to train/store/search/embed than the semantic vector search methods, and execution speed was a key consideration in this research, whereas other resource limitations (e.g., memory) were not.
Interestingly, the med2vec model which was trained on medical vocabulary did not perform particularly well during classification despite other researchers praising its ability on medical tasks. 
This may indicate that the structure of our documents was more significant than particular words or phrases.

\subsection{An unbalanced class distribution greatly biases the predictive accuracy of vector search classification}
The results suggest that the documents in class E were often identified correctly due to a numerical advantage, rather than their content.
Tab.~\ref{TAB:class_statistics} shows that many documents were classified as E, and the performance metrics in Fig.~\ref{FIG:classification_statistics} shows that these were easier to classify than documents which belonged to the classes A, B, C, or D.
Fig.~\ref{FIG:t-sne_visualisations} shows that many embeddings corresponding to this class were in close vector-space proximity to the other classes, and this corresponds to the results in Fig.~\ref{FIG:knn_and_predictive_performance} which show that increasing the number of nearest neighbours that were considered during classification was ineffective.
These results suggest that it was convenient for the vector search classification to classify documents as E if unsure. %

\subsection{More data is good}
Fig.~\ref{FIG:data_splits_and_predictive_performance} suggests that access to more data would increase predictive performance. %
This has been known by AI researchers and practitioners for a long time, and being able to access more medical records would be beneficial here too. 
Fig.~\ref{FIG:execution_times} indicates that much less time would be required to perform lexical vector search opposed to semantic vector search, but Tab.~\ref{TAB:embedding_dimensions} shows that TF and TF-IDF could suffer memory problems.

\section{Conclusion}
Our task was to classify medical documents according to their text. 
We used vector search to achieve this due to its simplicity, efficiency, significance, and many recent improvements.
Our results suggest that classifying rigidly-structured medical documents would be most successful if there were several documents of each target class, and a balanced number of document classes.
Additionally, %
we found that lexical vector search was slightly better at classifying our medical documents than off-the-shelf semantic vector search, but the embeddings were larger, and these required more memory and data examples.
This suggests that traditional machine learning techniques may, at present, outperform modern deep learning techniques in some situations. 
This does not mean that traditional machine learning is the best solution for every task, but these results indicate that it deserves to be considered. 
In the future, we would like to explore the problem of data drift, and at what point increasing the amount of data becomes detrimental.

\section{Acknowledgements}
We would like to thank all of colleagues from TMLEP and the University of Kent in Canterbury who were involved in this research.
This research was conducted as part of the ``medical data pagination'' knowledge transfer partnership project (grant number 10048265) that is funded by Innovate UK.

\bibliography{bib}

\end{document}